\documentclass[3p, twocolumn]{elsarticle}
\usepackage{amssymb}
\usepackage{amsmath}
\usepackage{graphics}
\usepackage{graphicx}
\usepackage[hyperindex=true,final]{hyperref}

\begin{document}

\title{Non-diffusive dynamics in a colloidal glass: aging versus rejuvenation}

\author[ISC]{R.~Angelini\corref{cor1}}
\author[ISC]{B.~Ruzicka}

\cortext[cor1]{Corresponding author email:roberta.angelini@roma1.infn.it.}

\address[ISC]{ISC-CNR and Dipartimento di Fisica, Sapienza Universit$\grave{a}$ di Roma, I-00185 Roma, Italy}

\date{\today}

\begin{abstract}

The microscopic dynamics of spontaneously aged and rejuvenated
glassy Laponite is investigated through X-ray photon correlation
spectroscopy. Two different behaviours of the intensity
autocorrelation functions are observed depending on the history of
the sample: stretched for spontaneously aged samples and samples
rejuvenated from a Wigner glass and compressed, typical of
anomalous dynamics, for samples rejuvenated from a DHOC glass. The
relaxation time behaviour in the three cases indicates a
non-diffusive dynamics of the particles. The present system offers
therefore an overview of various dynamical behaviours previously observed
individually in several systems  and the
possibility to pass from one to the other choosing ad hoc
the time parameter.

\end{abstract}

\maketitle


\vskip 10cm

\noindent\textbf{1. Introduction}

\begin{figure*}[t!]
\centering
\includegraphics[width=16cm,angle=0,clip]{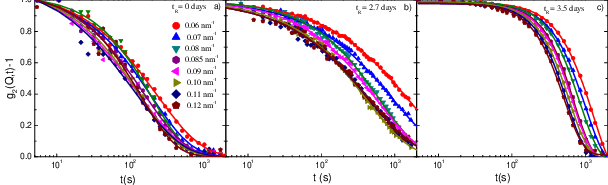}
\caption{Normalized intensity autocorrelation functions of an
aqueous Laponite  suspension at concentration C$_w$=3.0 $\% $  a)
for a spontaneously aged sample with $t_R$ = 0 days
($t_w$=5.67$\cdot10^4$ s), b) for a rejuvenated sample with $t_R$
=2.7 days ($t_w$=1.84$\cdot10^4$ s) and  c) for a rejuvenated
sample with $t_R$= 3.5 days ($t_w$=1.97$\cdot10^4$ s) and at
different Q values obtained through XPCS (symbols).  The solid
lines represent the best fits performed using Eq.~\ref{eq1}.}
\label{Fig1}
\end{figure*}

Soft matter plays nowadays a central role in the development of
advanced materials thanks to the variety of physical states which
offers including liquids, foams, gels, glasses. Its interest
arises from the ability that its microscopic constituents have to
rearrange and form mesoscopic structures. It is challenging
therefore to predict as much as possible these interesting
behaviours from the microscopic to the mesoscopic scales in order
to be able to manipulate and consequently control the macroscopic
behaviour of such materials. To this purpose the experimental and
theoretical study of the structure and dynamics of these systems
is essential. Studying the static and dynamic structure factors is
the most direct way to access information on the characteristic
length (dimensions) and time (relaxation times) scales typical of
these systems. Their dynamical behaviour is characterized by the
presence of relaxation processes~\cite{Balucani} associated for
example to the interactions between a particle and the cage of its
nearest neighbors or related to the structural rearrangements of
the particles. The signature of a relaxation process is a decay of
the dynamic structure factor described, over a wide time window,
by the Kohlrausch-Williams-Watts expression $f(Q,t)\sim
exp[-(t/\tau)^{\beta}]$ where $\tau$ is an ``effective''
relaxation time and $\beta$ measures the distribution of
relaxation times. Usually in soft materials an exponent $\beta<1$
is found, this behaviour is referred to as ``stretched
behaviour''. On the contrary in the last decade an unusual
behaviour characterized by a $\beta$ exponent $\beta>1$ has been
observed and referred to as ``compressed behaviour''~\cite{CipellettiPRL2000,BellourPRE2003,Bandyopadhyay_PRL_2004,
SchosselerPRE2006, FalusPRL2006,NarayananPRL2007,CaronnaPRL2008,
GuoPRL2009, DuriPRL2009,RutaPRL2012,OrsiPRL2012, AngeliniSM2013,
RutaSM2014,AngeliniNC2014,CristofoliniCOCIS2014}. These two
peculiar trends correspond to different microscopic dynamics of
the particles.

Here we present an overview of the dynamical behaviour of a
prototype charged colloidal system, aqueous Laponite suspensions,
characterized by an aging dynamics. Once dispersed in water
Laponite ages towards different arrested states like equilibrium
gel and glass depending on ionic strength and particle
concentrations~\cite{RuzickaSM2011}. In this work we focus on the
glass state at concentration $C_w=3.0$ $\%$ largely investigated
experimentally through different complementary techniques such as
Dynamic Light
Scattering~\cite{BellourPRE2003,RuzickaPRL2004,SchosselerPRE2006,Jabbari_PRE_2008},
Rheology~\cite{Joshi_RSPA_2008,Shahin_Langmuir_2010,Shahin_PRL_2011},
Small Angle X-ray Scattering~\cite{RuzickaPRE2008,RuzickaPRL2010,RuzickaNatMat2011}, Small
Angle Neutron Scattering~\cite{TudiscaPRE2014}, X-ray Photon
Correlation Spectroscopy~\cite{Bandyopadhyay_PRL_2004,AngeliniSM2013,AngeliniNC2014,MarquesSM2015} and Neutron Spin
Echo~\cite{MarquesSM2015}. The waiting time ($t_w$) (time scale)
and Q (spatial scale) dependence of the structural relaxation process of spontaneously aged and rejuvenated samples are studied. Both stretched and compressed behaviours of the intensity
autocorrelation functions are found and discussed in relation to
previous studies and on the light of the presence of attractive
and repulsive microscopic interactions versus the application of
external stresses. The different dynamical behaviours observed
here in a single system are general features individually observed
in several systems as molecular liquids~\cite{SciPRE1996,RinaldiPRE2001}, colloidal hard
spheres~\cite{SaltzmanPRE2006}, colloids~\cite{CipellettiPRL2000,DuriPRL2009},
clays~\cite{BellourPRE2003,Bandyopadhyay_PRL_2004,SchosselerPRE2006, AngeliniSM2013,
AngeliniNC2014}, metallic glasses~\cite{RutaPRL2012}, polymers~\cite{FalusPRL2006,NarayananPRL2007,GuoPRL2009,RutaSM2014},
supercooled liquids~\cite{CaronnaPRL2008}.

\noindent\textbf{2. Materials and Methods}

\noindent\textit{2.1 Materials}

Laponite is a synthetic clay that, when dispersed in water, forms
a charged colloidal suspension of platelets with 25 nm diameter
and 0.9 nm thickness and inhomogeneous charge distribution,
negative on the faces and positive on the rims. The platelets are
usually considered monodisperse in size but a small polydispersity
has been reported by different
authors~\cite{Kroon_PRE_1996,Balnois_LANG_2003}.

Aqueous dispersions of Laponite RD with weight concentrations
C$_w=3.0$ $\%$  were prepared 
in a glovebox under $N_2$ flux to avoid contact with air and prevent CO$_2$
degradation~\cite{Thompson_JCIS_1992}. Laponite powder,
manufactured by Laporte Ltd., was dispersed in pure deionized
water, stirred vigorously for 30 min and filtered soon thereafter
through a 0.45 $\mu$m pore size Millipore filter in bottles sealed
in the glovebox. A part of the stock solution was directly filtered
in glass capillaries of 2 mm diameter for the experiments. These
were later referred to as ``spontaneously aged" samples. The origin
of the waiting time ($t_w=0$) determines the age of the sample and
for the spontaneously aged samples it is the time at which the
suspension is filtered. Rejuvenated samples were prepared starting
from the stock solution that had rested in the bottles some time
$t_R$ (rejuvenation time) since filtration and injected into the
capillary through a syringe, hence introducing a huge shear field
(shear rejuvenation). The age $t_w$ of a rejuvenated sample is
counted from $t_R$. Even if rejuvenation by a shear field 
corresponds to returning the sample to earlier aging times, however the
process never rewinds the sample to the original as
prepared one.

\noindent\textit{2.2 Measurements}

The samples were characterized by X-ray Photon Correlation
Spectroscopy (XPCS) \cite{MadsenNJP2010,Leheny_COCIS_2012} at
beamline ID10A of the European Synchrotron Radiation Facility
(ESRF) in Grenoble. For the measurements a partially coherent and
monochromatic X-ray beam with a photon energy of 8 keV was
employed. Long series of scattering images were recorded by a
charged coupled device (CCD) placed in the forward scattering
direction. The images were post processed following the
multi-speckle XPCS approach \cite{MadsenNJP2010} to get access to
the dynamics of the samples. Ensemble averaged intensity
autocorrelation functions $g_2(Q,t)=\langle \frac{\langle
I(Q,t_0)I(Q,t_0+t)\rangle_p}{\langle I(Q,t_0)\rangle_p \langle
I(Q,t_0+t)\rangle_p }\rangle_{t_0}$ were calculated using a
standard multi-tau algorithm. Here, $\langle...\rangle_p$
indicates averaging over pixels of the detector mapping onto a
single value of the momentum transfer ($Q$) while
$\langle...\rangle_{t_0}$ indicates temporal averaging over $t_0$.

\noindent\textbf{3. Results and Discussion}

The XPCS intensity autocorrelation functions of aqueous Laponite
suspensions as a function of $Q$ are shown in Fig.~\ref{Fig1}
for three different samples: spontaneously aged
(Fig.~\ref{Fig1}a) rejuvenated before 3 days (Fig.~\ref{Fig1}b)
and rejuvenated after 3 days (Fig.~\ref{Fig1}c). The fits through the expression

\begin{equation}\label{eq1}
g_2(Q,t)-1=(C\exp(-(t/\tau)^{\beta}))^2
\end{equation}

are reported  as full lines in Fig.~\ref{Fig1}. Here $C^2$ represents the contrast,  $\tau$ the structural
relaxation time and $\beta$ the Kohlrausch exponent which
characterize the dynamics of the sample.
Two behaviours 
can be  distinguished: stretched ($\beta<1$) for the spontaneously aged (Fig.~\ref{Fig1}a) and the rejuvenated before 3 days (Fig.~\ref{Fig1}b) samples
and compressed ($\beta>1$) for the sample rejuvenated after 3 days
(Fig.~\ref{Fig1}c). The distinct dynamical behaviours of the rejuvenated samples are
related to  a recently discovered glass-glass transition which
spontaneously happens 
between a repulsive glass (Wigner glass) and a Disconnected House
of Cards Glass (DHOC)  stabilized by attractive interactions that
partially orient the particles~\cite{AngeliniNC2014}. It happens at a critical time $t_c \approx$ 3 days
and different rejuvenated states are obtained if the rejuvenation
process is applied to a Wigner glass or to a DHOC glass, i.e. if
 $t_R<t_c$ or $t_R>t_c$ respectively. 
These results are in agreement with rheological measurements on rejuvenated samples that showed how dynamical properties of these systems strongly depend on the time elapsed between preparation and rejuvenation~\cite{Shahin_Langmuir_2010}. Furthermore in another recent study on the effect of shear melting and rejuvenation on the aging of colloidal suspensions of Laponite at $C_w=2.8$ \% with the addition of salt~\cite{JatavJR2014}, a different rheological behaviour between freshly prepared and rejuvenated samples is found. This difference is attributed to the inability of the shear melting to completely break the arrested structure in the rejuvenated samples thereby creating unbroken aggregates with consequent different rheological response between spontaneously aged and rejuvenated samples. Most importantly the slope of the relaxation time spectra shows two distinct rheological responses for the rejuvenated samples, passing from negative for small rest times (equivalent to our rejuvenation time) (0.5 days, 1 day, 1.2 days) to positive for high rest times (6 days). Although the sample has a slight lower concentration ($C_w=2.8$ \%), which implies a slightly slower dynamics with respect to our sample ($C_w=3.0$ \%) and it is prepared with added  salt (0.3 mM), which fasten the aging dynamics, the phenomenology is surprisingly in agreement with our findings, confirming the existence of three rheological behaviours for spontaneously aged samples and samples rejuvenated before and after a certain critical time." 

\begin{figure}[t!]
\centering
\includegraphics[width=7.5cm,angle=0,clip]{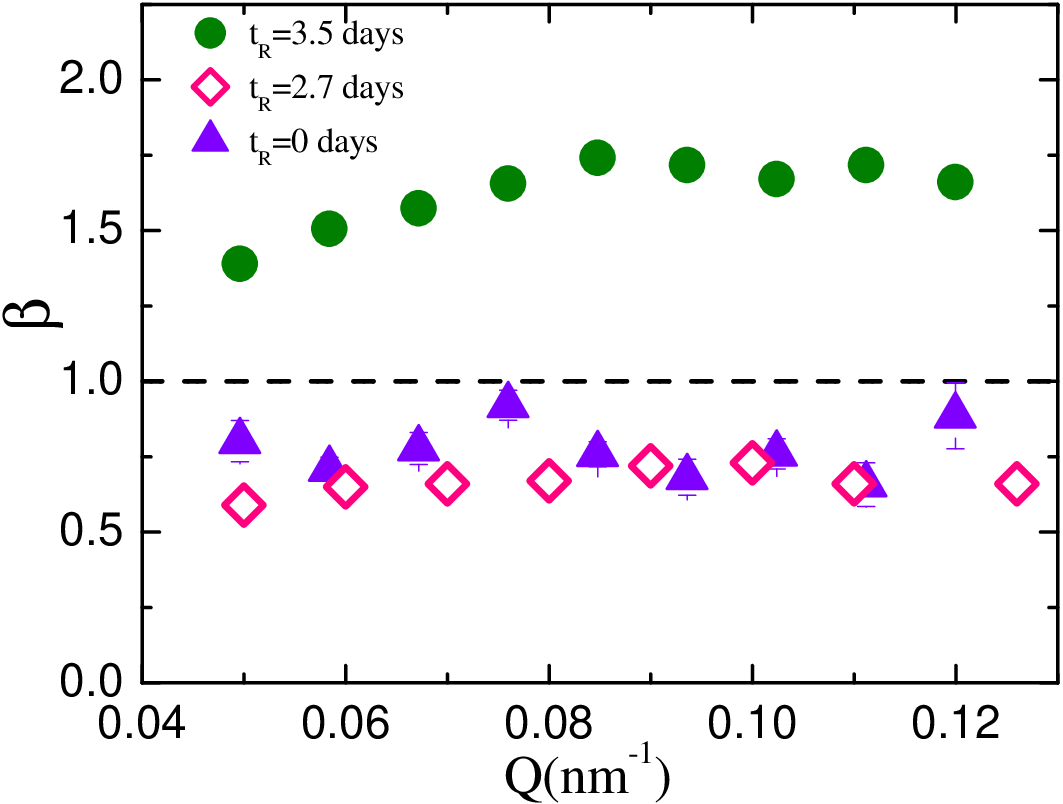}
\caption{Kohlrausch exponent $\beta$ as a function of Q obtained
from Eq.~\ref{eq1} for an aqueous Laponite suspension at
concentration  C$_w$=3.0 $\% $  spontaneously aged with $t_R$ = 0
days ($\beta<1$), rejuvenated  with $t_R$ =2.7 days ($\beta<1$)
and  rejuvenated with $t_R$= 3.5 days ($\beta>1$).} \label{Fig2}
\end{figure}

To quantify the qualitative difference observed above, the exponent
$\beta$ obtained by Eq.~\ref{eq1} is plotted vs $Q$ in
Fig.~\ref{Fig2} for the spontaneously aged and the rejuvenated
samples.  The value of $\beta$ is well below 1 for the
spontaneously aged sample (closed triangles)
and for sample rejuvenated  before $t_c\approx$ 3 days (open
diamond)~\cite{AngeliniNC2014}. This indicates that the
correlation functions have stretched exponential shape
(Fig.~\ref{Fig1}a,b,) as commonly observed for glass dynamics. On
the contrary $\beta$ is always above unity for the sample
rejuvenated after $t_c\approx$ 3 days (closed
circles)~\cite{AngeliniNC2014}, implying a compressed exponential
shape of the correlation curves (Fig.~\ref{Fig1}c),
signature of an anomalous dynamics.

\begin{figure}[t]
\centering
\includegraphics[width=7.5cm,angle=0,clip]{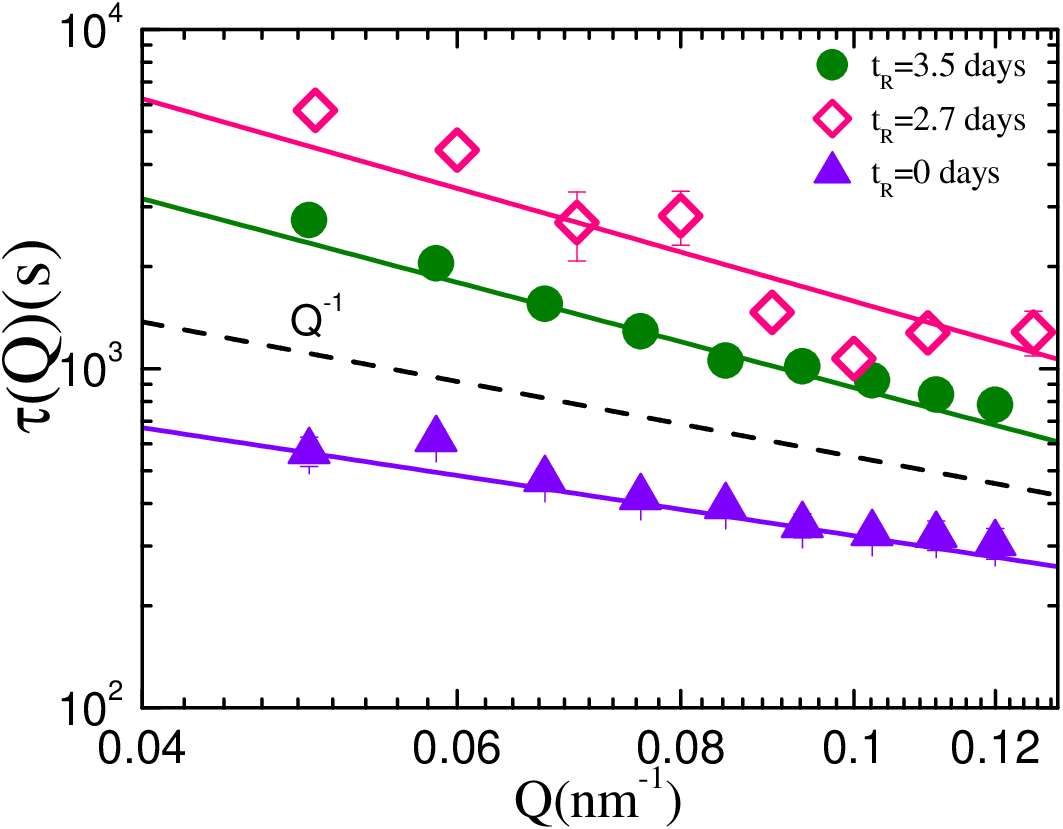}
\caption{Structural relaxation time as a function of Q obtained
from Eq.~\ref{eq1} for an aqueous Laponite suspension at
concentration  C$_w$=3.0 $\% $ spontaneously aged ($t_R$ = 0
days),  rejuvenated with $t_R$ =2.7 days and rejuvenated with
$t_R$= 3.5 days. The full lines represent fits of $\tau(Q)$ with a
power law. } \label{Fig3}
\end{figure}

To better understand the dynamical behaviour of the three samples
the $Q$ dependence of the structural relaxation time, as obtained
through the fit with Eq.~\ref{eq1}, is reported in
Fig.~\ref{Fig3}. At variance with the $\beta$ exponent the
behaviour of the relaxation time is very similar for the three
cases with  $\tau(Q) \sim Q^{-1}$, signature of non-free diffusive
dynamics.

\begin{figure}[t]
\centering
\includegraphics[width=7.8cm,angle=0,clip]{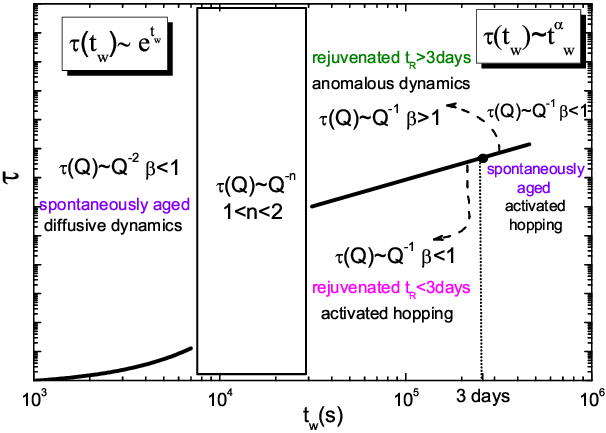}
\caption{Structural relaxation time as a function of waiting time
 together with a summary of the
different  observed dynamical behaviours.} \label{Fig4}
\end{figure}

The different dynamical behaviours observed and their dependence
on waiting time, $Q$ and history are summarized in
Fig.~\ref{Fig4}. Previous studies agree on the existence of a
typical dual waiting time dependence of the structural relaxation
time $\tau$~\cite{CipellettiPRL2000,
BellourPRE2003,TanakaPRE2005,SchosselerPRE2006,AngeliniSM2013,
MarquesSM2015}: an exponential waiting time dependence $\tau(t_w)
\sim e^{t_w}$ at small $t_w$, in the cage forming
regime~\cite{TanakaPRE2005}, and a waiting time power law
dependence $\tau(t_w) \sim t_w^{\alpha}$ with
 $\alpha \sim$ 1 at larger
$t_w$, in the full aging regime~\cite{TanakaPRE2005}. While in the
cage forming regime the dynamical behaviour is diffusive
($\tau(Q)\sim Q^{-2}$) with stretched ($\beta<$1) intensity
correlation functions, in the full aging regime the dynamics is no
more diffusive but characterized by a $\tau(Q)\sim Q^{-1}$
dependence with three different behaviours of the correlation
functions depending on the history of the sample. In particular
both the sample spontaneously aged and rejuvenated before $t_c
\approx$ 3 days have stretched exponential correlation curves
($\beta<1$). The only theoretical model explaining this dynamical
behaviour, attributed to activated hopping, is formulated by Bhattacharyya et al.~\cite{BhattacharyyaJCP2010}. In the case of samples
rejuvenated after $t_c \approx$ 3 days  the scenario is different
and $\tau(Q) \sim Q^{-1}$ and $\beta>1$ are found, this is the
case of anomalous dynamics as reported in a big variety of
different systems~\cite{CipellettiPRL2000,BellourPRE2003,
Bandyopadhyay_PRL_2004,SchosselerPRE2006,FalusPRL2006,CaronnaPRL2008,GuoPRL2009,RutaPRL2012,RutaSM2014}.
It has been attributed to the relaxation of internal stresses~\cite{CipellettiPRL2000, CipellettiFD2003,BouchaudEPJE2001}. 
In between the cage forming and full aging regimes MD
simulations~\cite{MarquesSM2015} find, in spontaneously aged
sample, stretched exponential curves and a relaxation time well
described as $\tau(Q) \sim Q^{-n}$ with $1<n<2$.

The comparison between the three samples here reported shows how
in a single colloidal suspension different dynamical behaviours
can be found and how it is possible to trigger the system from a
behaviour to the other simply waiting time and/or changing the
time elapsed before rejuvenation. We want to stress that even if
the peculiar and complex nature of aqueous Laponite
suspensions~\cite{RuzickaSM2011} renders possible the presence of
different behaviours in a single system, the same individual
behaviours are found separately in different materials. In
particular, on one side non-diffusive/activated dynamics ($\tau(Q)
\sim Q^{-1}$, $\beta<1$) has been detected in the supercooled
regime of molecular liquids (such as water~\cite{SciPRE1996} and
ortho-terphenyle~\cite{RinaldiPRE2001}) and of colloidal hard
spheres~\cite{SaltzmanPRE2006}. On the other side the anomalous
dynamics ($\tau(Q) \sim Q^{-1}$, $\beta>1$) is today considered a
general feature of several systems as
colloids~\cite{CipellettiPRL2000,DuriPRL2009},
clays~\cite{BellourPRE2003,
Bandyopadhyay_PRL_2004,SchosselerPRE2006, AngeliniSM2013,
AngeliniNC2014}, metallic glasses~\cite{RutaPRL2012},
polymers~\cite{FalusPRL2006,NarayananPRL2007,GuoPRL2009,RutaSM2014},
supercooled liquids~\cite{CaronnaPRL2008}, etc. However despite
the big amount of experimental evidences a theoretical model that
explains these behaviours is not available and the exact
understanding of the origin of these two different dynamical
behaviours is still lacking.

\noindent\textbf{4. Conclusions}

In conclusion, we have reported an overview of recent findings on
the dynamical behaviour of a colloidal glass. Three different
cases have been compared: a spontaneously aged sample, a sample
rejuvenated from a Wigner glass and a sample rejuvenated from a
DHOC glass. A long time non-diffusive dynamics ($\tau(Q) \sim
Q^{-1}$) has been found. It is associated to different decays of
the intensity correlation functions: stretched ($\beta<1$) for
spontaneously aged and rejuvenated Wigner glass samples and
compressed ($\beta>1$) for rejuvenated DHOC glass sample. The
parameter (time) that allows to pass from one behaviour to the
other has been identified and can be chosen ad hoc to control the
dynamics of the system.

\noindent\textbf{Acknowledgments}

We acknowledge ESRF for beamtime for this project. The staff at
beamline ID10A is acknowledged for help during the experiments. RA
acknowledges support from MIUR-PRIN.


\end{document}